\documentclass{iopart}
\usepackage{graphicx,epsfig}
\usepackage{bm}
\usepackage{iopams}
\newcommand{\ba}{\begin{eqnarray}}
\newcommand{\ea}{\end{eqnarray}}
\newcommand{\bse}{\numparts}
\newcommand{\ese}{\endnumparts}

\newcommand{\B}{{\cal{B}}}

\newcommand{\C}{{\cal {C}}}
\newcommand{\DD}{{\cal {D}}}

\newcommand{\W}{{\cal {W}}}
\newcommand{\bbq}{\begin{quote}}
\newcommand{\eeq}{\end{quote}}
\newcommand{\RR}{{}^3{\cal{R}}}

\newcommand{\T}{{}^3{\cal{T}}}

\newcommand{\EE}{{\cal{E}}}
\newcommand{\FF}{{\cal{F}}}
\newcommand{\JJ}{{\cal{J}}}
\newcommand{\VV}{{\cal{V}}}
\newcommand{\HH}{{\cal{H}}}

\newcommand{\PP}{{\cal{P}}}
\newcommand{\MM}{{\cal{M}}}
\newcommand{\QQ}{{\cal{Q}}}

\newcommand{\rhoav}{\langle\rho\rangle}

\newcommand{\muav}{\langle\mu\rangle}

\newcommand{\HHav}{\langle\HH\rangle}

\newcommand{\Thetaav}{\langle\Theta\rangle}

\newcommand{\RRav}{\langle\RR\rangle}
\newcommand{\pav}{\langle p\rangle}

\newcommand{\FFav}{\langle\FF\rangle}

\newcommand{\Aav}{\langle A\rangle}
\newcommand{\Bav}{\langle B\rangle}

\newcommand{\Da}{\delta^{(A)}}

\newcommand{\Dth}{\delta^{(\Theta)}}
\newcommand{\dDth}{\dot\delta^{(\Theta)}}
\newcommand{\Dm}{\delta^{(\mu)}}
\newcommand{\dDm}{\dot\delta^{(\mu)}}

\newcommand{\Dp}{\delta^{(p)}}

\newcommand{\dRR}{{}^3\dot{\cal{R}}}
\newcommand{\DRR}{\delta^{(\RR)}}

\begin{document}


\title[Quasi--local variables, non--linear perturbations and back--reaction]{Quasi--local variables, non--linear perturbations and back--reaction in spherically symmetric spacetimes.} 
\author{ Roberto A. Sussman}
\address{Instituto de Ciencias Nucleares, Universidad Nacional Aut\'onoma de M\'exico (ICN-UNAM),
A. P. 70--543, 04510 M\'exico D. F., M\'exico. }
\ead{sussman@nucleares.unam.mx}
\date{\today}
\begin{abstract} We introduce a quasi--local integral functional and scalar quasi-local variables to examine a wide class of spherically symmetric inhomogeneous spacetimes that generalize the Lema\^\i tre-Tolman-Bondi (LTB) dust solutions (``LTB'' spacetimes). By using these variables, we can transform the fluid flow evolution equations into evolution equations for non-linear, covariant, gauge--invariant perturbations of Friedman-Lema\^\i tre-Robertson-Walker (FLRW) cosmologies. In the linear limit, we obtain spherical perturbations in the synchronous gauge under the long wavelength approximation. The formalism has a significant potential for cosmological applications, as it allows one to examine a wide variety of sources with different ``equations of state'', generalizing known FLRW solutions to idealized but non--trivial and non--linear inhomogeneous conditions. The quasi--local functional can be reformulated as a weighed proper volume average distribution, with the weight factor given by a scalar invariant related to the quasi--local mass--energy function. The back--reaction terms, emerging in Buchert's proper averaging formalism, can be expressed as  differences between fluctuations of averaged and quasi-local energy densities. By comparing this average with the weighed quasi-local one, we can define a binding energy functional related to spatial gradients of the averaged and quasi--local variables that appear in the back--reaction terms.             
\end{abstract}

\section{Introduction.}
 
Recent observations seem to favor a dominant theoretical consensus, ``the concordance model', in which present cosmic dynamics is dominated by an elusive source called ``dark energy'', somehow associated with a repulsive (yet unknown) interaction\cite{review}. While the other dominant cosmic source, dark matter, is taken as inhomogeneous due to its association with structure formation at the galactic scale, dark energy is mostly considered as homogeneous, or nearly so, as it is assumed to have a signifficant effect only at a larger cosmic scale in which the universe appears to be homogeneous. However, given our ignorance on the fundamental properties of these sources, there is no theoretically binding reason to assume, {\it a priori}, that no valuable new information would result from studying their interaction under inhomogeneous conditions, at least in scales associated with structure formation and gravitational clustering. 

Since non--linear inhomogeneity in gravitational clustering or virialization of bound structure occurs at Newtonian velocities, these phenomena have been studied in a Newtonian framework~\cite{NewtDE1,NewtDE2,NewtDE3,NewtDE4}. However, there is no reason to avoid using the low--energy regime of a relativistic theory, rather than a pure Newtonian framework. After all, rest--mass is the only source of Newtonian gravity, so it is unreasonable to assume that the main effect of these unknown sources can be reduced to rest--mass, ignoring possible effects from (yet unknown) forms of matter--energy contributions and their interactions. In fact, even with a cosmological constant, the study of the equilibrium and stability of astrophysical sources yields different results in the Newtonian limit of General Relativity from those of a pure Newtonian approach \cite{LNewt}.

A large body of recent research considers a relativistic treatment of inhomogeneous sources in the galactic and cosmic scale as part of the critique of the concordance model. For example,  the possibility that cosmic acceleration (or at least part of it) could be an effect of due to the presence of inhomogeneites in photon trajectories in observations from high red--shift objects~\cite{InhObs1, InhObs2,InhObs3,InhObs4,InhObs5,InhObs6,InhObs7}. From a more theoretical angle, a number of articles relate or explain cosmic acceleration in connection with non--local effects that appear in various averaging formalisms~\cite{BRA1,BRA2,BRA3,BRA4,BRA5,BRA6}. In particular, an  ``effective'' acceleration  emerges in connection with the so--called ``back--reaction'' in Buchert's scalar average formalism \cite{buchert}. This acceleration can be negative and could be, in principle, observed~\cite{obs_BR1,obs_BR2,test_BR}. Also, back--reaction can be associated with the existence of non--trivial gradients of quasi--local energy between gravitationally bound systems in an expanding cosmic background.~\cite{BRA4,wilt1,wilt2} 

While the aim of most of the above mentioned proposals is to question the dark energy paradigm, they could also be compatible with it, as their predicted effects could provide just a partial explanation for the observed acceleration. However, even if the dark energy explanation prevails over the alternatives that dispute it, a controversy which is beyond the scope of this article, there is no harm done (and perhaps valuable new information) in probing its behavior under inhomogeneous conditions. As a compromise between ``realistic'' inhomogeneity that would require numerical methods and codes of high complexity and the approach of linear perturbations, we examine in this article inhomogeneous spacetimes under idealized but non--trivial conditions. We describe and summarize below the contents of the article.

By assuming a spherically symmetric Lema\^\i tre--Tolman--Bondi metric, we derive in section \ref{genLTB} a class of spacetimes (``LTB spacetimes'') that generalize to nonzero pressure (and nonzero pressure gradients) the well know solutions for inhomogeneous dust associated with this metric \cite{kras}. The most general source for this metric (in the comoving frame) is a fluid with anisotropic stresses, which will end up being interpreted as fluctuations of the isotropic pressure. However, regardless of the interpretation, we remark that the existence of pressure anisotropy is far from a drawback or shortcoming, not only because it can be associated to a number of well motivated physical effects~\cite{anisP}, but because an inhomogenous source with only isotropic pressure is a far more idealized situation than one with anisotropic pressure.   

In section \ref{1plus3} we show that LTB spacetimes can be fully characterized by a set of covariant scalars, whose evolution equations in the covariant fluid flow ``1+3'' representation~\cite{ellisbruni89,BDE,1plus3} reduce to a system of non--linear autonomous scalar PDE's and constraints involving radial gradients. In section \ref{QLdefs}, we introduce a ``quasi--local'' integral functional, and its associated ``quasi--local'' scalar functions, related to the Misner--Sharp quasi--local mass--energy~\cite{MSQLM,szab,hayward1,hayward2}. Expressing all local covariant scalars as perturbations of the quasi--local scalars, we rewrite in section \ref{eveqs} the ``1+3'' fluid flow equations as evolution equations for the quasi--local variables and their perturbations.  These equations are effectively equivalent to a system of ODE's in which radial dependency enters as a parameter (see Appendix C of \cite{suss08}). 

We prove in section \ref{perturb}, on the basis of known criteria that define a perturbation formalism~\cite{ellisbruni89, BDE,1plus3,bardeen}, that the quasi--local evolution equations can be rigorously characterized as  evolution equations of covariant, gauge--invariant, non--linear perturbations on a FLRW formal ``background'' given by the quasi--local variables (which satisfy FLRW dynamical equations). All scalars associated with quantities that vanish in the FLRW background: Weyl tensor, anisotropic pressure and shear are expressible as gauge invariant first order fluctuations of the energy density, isotropic pressure and expansion scalar around their corresponding quasi--local variables.

As we show in section \ref{EOS}, once we choose an ``equation of state'' (EOS) between the quasi--local energy density and pressure (which are ``background'' variables), the whole system of evolution equations becomes determined. Given this equation of state, we point out that non--local effects associated with the long range nature of the interactions in self--gravitating sources~\cite{Padma_GTD} can arise when comparing local and quasi--local scalars. In order to examine this issue, we look in section \ref{idealgas} at the classical ideal gas, showing that non--linear perturbations of particle number density and temperature provide a first order ``virial'' correction (given as a correlation energy) to the ``usual'' EOS satisfied by the background variables. This type of virial corrections to the ideal gas EOS have been found in Newtonian self--gravitating systems~\cite{saslaw1,saslaw2,saslaw3}. Similar effects could arise in applications of the perturbation formalism to self--gravitating relativistic dark energy sources whose fundamental physics is still unknown. 

We verify in section \ref{linear} that in the linear limit, when perturbations are ``small'', these evolution equations reduce to those of spherical linear perturbations in the synchronous gauge, but under the long wave length approximation~\cite{padma2,hwangnoh}. In section \ref{covariant} we rewrite, in full covariant form, the quasi--local functional and scalar functions in terms of weighed proper volume integrals, where the ``weight factor'' is an invariant scalar associated with the quasi--local mass--energy. The quasi--local functional becomes then a weighed spatial average distribution that generalizes the standard spatial average. By comparing in section \ref{QLaverage} the integral mass--energy functionals associated with the quasi--local (weighed) and standard (weight = 1) averages, we can define a domain--dependent binding energy.  

A theoretical connection between quasi--local variables and the issue of back--reaction has been suggested by Wiltshire~\cite{BRA4,wilt1,wilt2}, in the context of the scalar averaging formalism of Buchert and coworkers. We provide useful arguments to this suggestion in section \ref{BR}, by showing that both kinematic and dynamical back--reaction terms are essentially comparisons of the square fluctuations of the expansion scalar and pressure around their spatial average and around their quasi--local equivalents. As shown elsewhere~\cite{sussBR}, this comparison provides sufficient conditions for a positive back--reaction (and necessary for a negative ``effective'' acceleration). These  fluctuations can be expressed in terms of spatial gradients of the averaged and quasi--local variables, and also in terms of the domain dependent behavior of the binding energy. In section \ref{conclusion} we provide a brief summary of our results and suggest possible avenues for further research. We discuss several regularity issues in the Appendix.

In follow up articles we will examine numeric applications of the formalism presented here, looking at the issue of back--reaction, as well as the modelling of cosmological dark matter and dark energy sources.

\section{Generalized LTB spacetimes}\label{genLTB}

We will use the term ``LTB spacetimes'' to denote all solutions of Einstein's equations described by the spherically symmetric Lema\^\i tre--Tolman--Bondi metric in a comoving frame:
\begin{equation} ds^2 = -c^2dt^2 +\frac{R'^2\,dr^2}{1-K}+R^2[d\theta^2+\sin^2\theta\,d\phi^2],\label{LTB}\end{equation}
where $R=R(ct,r)$, \, $R'=\partial R/\partial r$ and $K=K(r)$. LTB spacetimes comply with: (a) 4--velocity is a geodesic field, (b) zero energy flow in a comoving frame and (c) existence of a regular symmetry center~\cite{kras}.  The metric (\ref{LTB}) is usually associated with a dust source, but the most general source compatible with it is an anisotropic fluid
\begin{equation}T^{ab} = \mu\,u^a u^b + p\,h^{ab}+\Pi^{ab},\label{Tab}\end{equation}  
where $u^a=\delta^a_0$,\,with $\mu(ct,r),\,p(ct,r)$ being, respectively, the energy density and isotropic pressure, $\Pi^{ab}$ is the anisotropic pressure trace-less tensor ($\Pi^a_a=u_a\Pi^{ab}=0$) and $h_{ab}=u_au_b+g_{ab}=\delta_a^i\delta_b^jg_{ij}$ is the induced metric of hypersurfaces $\T(t)$ orthogonal to the 4--velocity ($t$ constant, with space indexes $i,j=r,\theta,\phi$).

Besides $u^a,\,\mu,\,p$ and $\Pi^{ab}$, the remaining covariant objects associated with LTB spacetimes are
\begin{itemize}
\item Expansion scalar:
\begin{equation} \Theta = \nabla_a u^a = \frac{2\dot R}{R}+\frac{\dot R'}{R'},\label{theta1}\end{equation}
\item Ricci scalar of hypersurfaces $\T(t)$:
\begin{equation} \RR  = \frac{2(KR)'}{R^2R'},\label{RR1}\end{equation}
\item Shear tensor:
\begin{equation} \sigma_{ab}=\nabla_{(a}u_{a)}-\frac{\theta}{3}h_{ab},\label{sigma1}\end{equation}
\item The ``electric'' Weyl tensor:
\begin{equation} E^{ab} = u_c u_d C^{abcd},\label{EEtens}\end{equation}
\end{itemize}
where $\dot R=\partial R/\partial ct$ and $C^{abcd}$ is the Weyl curvature tensor.

In LTB spacetimes all spacelike symmetric trace--less tensors  like
$\Pi^{ab},\,\sigma^{ab},\,E^{ab}$ can be expressed in terms of a single scalar function as 
\bse\label{PSEsc}\ba \Pi^{ab} &=& \PP\,\Xi^{ab},\qquad\Rightarrow\quad \PP = \Xi_{ab}\Pi^{ab},\label{PPsc}\\ 
\sigma^{ab}&=&\Sigma\,\Xi^{ab},\qquad\Rightarrow\quad \Sigma = \Xi_{ab}\sigma^{ab},\label{Sigsc}\\ E^{ab}&=&
\EE\,\Xi^{ab},\qquad\Rightarrow\quad \EE = \Xi_{ab}E^{ab},\label{EEsc}\ea\ese
where $\Xi^{ab}\equiv h^{ab}-3\chi^{a}\chi^b$\, and $\chi^a=\sqrt{h^{rr}}\,\delta^a_r$ is the unit vector orthogonal to $u^a$ and to the 2-spheres, orbits of SO(3). 

The scalars $\mu,\,p,\,\PP$ follow from the field equations $G^{ab}=\kappa T^{ab}$ (with $\kappa =8\pi G/c^4$) for (\ref{LTB})--(\ref{Tab}):
\bse\label{fieldeqs}\ba \kappa\,\mu\,R^2R' &=& \left[R(\dot R^2+K)\right]',\label{mu1}\\
\kappa\,p\,R^2R' &=& -\frac{1}{3}\left[R(\dot R^2+K)+2R^2\ddot R\right]',\label{p1}\\
\kappa\,\PP\,\frac{R'}{R} &=& -\frac{1}{6}\left[\frac{\dot R^2+K}{R^2} +\frac{2\ddot
Y}{Y}\right]',\label{PP1}\ea\ese
while from (\ref{theta1}) and (\ref{PPsc})--(\ref{EEsc}), we obtain for $\EE$ and $\Sigma$ 
\bse\label{ESsc}\ba\Sigma &=& \frac{1}{3}\left[\frac{\dot R}{R}-\frac{\dot R'}{R'}\right],\label{Sigma1}\\
\EE &=& -\frac{\kappa}{2}\,\PP-\frac{\kappa}{6}\,\mu+ \frac{\dot
R^2+K}{2R^2}.\label{EE_1}\ea\ese
It is evident that LTB spacetimes are completely characterized by the following set of covariant scalars
\begin{equation} \{\mu,\,p,\,\PP,\,\Theta,\,\Sigma,\,\EE,\,\RR\},\label{loc_scals}\end{equation}
which will be called the ``local'' scalar representation. 

The best known LTB spacetime is that with a dust source: $p=\PP=0$, for which total energy density is rest--mass density: $\mu=\rho c^2$, while (\ref{mu1})--(\ref{PP1}) reduce to
\ba \kappa \rho c^2 R^2R' = 2M',\label{rho_dust}
\\ \dot R^2 = \frac{2M}{R}-K,\label{fried_dust}
\ea
where $M=M(r)$, which appears as an ``integration constant', is the ``effective'' rest mass--energy in length units. The rest of the scalars $\Theta,\,\Sigma,\, \RR,\, \EE$ remain the same (with $\PP=0$ in (\ref{EE_1})). The standard tactic to deal with the dust case is to solve the Friedman equation (\ref{fried_dust}) for a given choice of $M(r)$ and $K(r)$, and then to evaluate from this solution all the derivatives of $R,\,M,\,K$ in order to compute the scalars $\rho,\,\Theta,\,\Sigma,\, \RR,\, \EE$  from (\ref{theta1}), (\ref{RR1}), (\ref{Sigma1}), (\ref{EE_1}) and (\ref{rho_dust}). There is an extensive literature on how this is done~\cite{ltbstuff,suss02} (for an extensive review, see \cite{kras}), and the same tactic can work on simple particular cases with non--zero pressure and simple ``equations of state'' relating $\mu,\,p$ and $\PP$ (see \cite{kras,hydro,intmix3}), but for a general LTB spacetime it is not possible (or very difficult) to find (not to mention solve) an equation like (\ref{fried_dust}) in order to evaluate radial gradients of $R$ for computing the scalars in (\ref{loc_scals}). 

\section{The ``1+3'' decomposition}\label{1plus3}

An alternative approach to LTB spacetimes follows from the covariant ``1+3'' spacetime decomposition of Ehlers, Ellis, Dunsbury and van Ellst~\cite{ellisbruni89,BDE,1plus3}, or ``fluid flow'' equations, in which the field and conservation equations $\nabla_a T^{ab}=0$ for (\ref{LTB})--(\ref{Tab}) are transformed into a first order system of time evolution equations and constraints for the covariant scalars (\ref{loc_scals}). 

Bearing in mind (\ref{PPsc})--(\ref{EEsc}), the ``1+3'' equations for LTB spacetimes become the following set of scalar evolution equations
\bse\label{eveqs_13}\ba
\dot\Theta &=&-\frac{\Theta^2}{3}
-\frac{\kappa}{2}\left(\,\mu+3p\,\right)-6\Sigma^2,\label{ev_theta_13}\\
\dot \mu &=& -(\mu+p)\,\Theta-6\Sigma \PP,\label{ev_mu_13}\\
\dot\Sigma &=& -\frac{2\Theta}{3}\,\Sigma+\Sigma^2-\EE+\frac{\kappa}{2}\PP,
\label{ev_Sigma_13}\\
 \dot\EE &=& -\frac{\kappa}{2}\dot \PP-\frac{\kappa}{2}\left[\mu+p-2\PP\right]\Sigma
-3\left(\EE+\frac{\kappa}{6}\PP\right)\left(\frac{\Theta}{3}+\Sigma\right),\label{ev_EE_13}\ea\ese
together with the spacelike constraints  
\bse\label{c_13}\ba (p-2\PP)\,'-6\,\PP\,\frac{R'}{R}=0,\label{cPP_13}\\
\left(\Sigma+\frac{\Theta}{3}\right)'+3\,\Sigma\,\frac{R'}{R}=0,\label{cSigma_13}\\
\frac{\kappa}{6}\left(\mu+\frac{3}{2}\PP\right)'
+\EE\,'+3\,\EE\,\frac{R'}{R}=0,\label{cEE_13}\ea\ese
and the Friedman equation (or ``Hamiltonian'' constraint)
\begin{equation}\left(\frac{\Theta}{3}\right)^2 = \frac{\kappa}{3}\, \mu
-\frac{\RR}{6}+\Sigma^2,\label{cHam_13}\end{equation}
which is an integral of the Raychaudhuri equation (\ref{ev_theta_13}). The system
(\ref{ev_theta_13})--(\ref{cHam_13}) requires an equation of state (EOS) linking $\mu,\,p$ and
$\PP$ to become determined. However, time and radial derivatives do not decouple, in general.
Moreover, (\ref{loc_scals}) is not the only scalar representation for LTB spacetimes. 

\section{The quasi--local functional and scalar functions.} \label{QLdefs}

The Misner--Sharp quasi--local mass--energy function $\MM$ is a well known invariant defined for spherically symmetric spacetimes as \cite{MSQLM,szab,hayward1,hayward2}:
\ba \fl \frac{2\MM}{R} \equiv R^2{\cal{R}}^{(a)(b)}\,_{(a)(b)}{\vartheta}_{(a)}{\varphi}_{(b)}{\vartheta}^{(a)}{\varphi}^{(b)} =
1-\frac{(\partial_r R)^2}{g_{rr}}-\frac{(\partial_0 R)^2}{g_{00}},\label{MSQLM}\ea
where ${\cal{R}}^{(a)(b)}\,_{(c)(d)}$ is the Riemann tensor in an orthonormal tetrad and ${\vartheta}^{(a)}=\sqrt{h^{\theta\theta}}\delta^a_\theta$,\, ${\varphi}^{(b)}=\sqrt{h^{\phi\phi}}\delta^b_\phi$ are the tetrad unit vectors tangent to the orbits of SO(3). In fact, the two main invariants in spherically symmetric spacetimes are precisely $\MM$ and $R$ \cite{hayward1}, so that quantities expressible in terms of these invariants or their derivatives are necessarily covariant.  

For LTB spacetimes with an anisotropic source tensor like (\ref{Tab}), the definition (\ref{MSQLM}) leads to:
\bse\label{QLM_expr}\ba \frac{2\MM}{R} &=& \dot R^2 + K, \label{QLM_LTB}\\
 2\MM' &=& \kappa \mu R^2 R', \label{QLM1}\\
 2\dot\MM &=& -\kappa (p-2\PP) R^2 \dot R.\label{QLM2} \ea\ese
whose integrability condition is equivalent to the conservation law $\nabla_b T^{ab}=0$ (equations (\ref{ev_mu_13}) and (\ref{cPP_13})). Notice that for the particular case of dust ($p=\PP=0$) we have $\MM =M$ and the quasi--local mass--energy is conserved along the 4--velocity flow: $\dot \MM=0$. When $p,\,\PP$ are nonzero, then (\ref{QLM_LTB}) generalizes the Friedman equation (\ref{fried_dust}). 

By looking at (\ref{QLM_LTB}) and (\ref{QLM1}), and assuming the existence of a regular symmetry center at $r=0$, an integral expression for $\MM$ can be constructed by integrating both sides of (\ref{mu1}) along spherical comoving domains in hypersurfaces $\T(t)$:~\cite{hayward1}
\begin{equation} \frac{\kappa}{3}\frac{\int_{x=0}^{x=r}{\mu\,R^2R' {\rm d} x}}{\int_{x=0}^{x=r}{R^2R' {\rm d} x}}=\frac{\dot R^2+K}{R^2}.\label{QLM_mu}\end{equation}
Different forms of mass--energy cannot be split or localized (in a covariant covariant) in General Relativity, but the function $\MM$ obtained by means of (\ref{QLM_mu}) can be interpreted as a total or ``effective'' mass--energy associated with a compact domain bounded by the 2--sphere marked by $r$.  This function also has a well understood behavior in the appropriate asymptotic limits~\cite{MSQLM,szab,hayward1,hayward2}. The integral (\ref{QLM_mu}) motivates us to define an integral functional acting on scalar functions defined in this type of compact domains:\\

\begin{quote}

\noindent
{\underline{Definition 1. Quasi--local functional.}} Let $X(\DD)$ be the set of all smooth integrable scalar functions in the regular comoving domain $\DD=\mathbb{S}^2\times\xi_r\subset \T(t)$, where $\mathbb{S}^2$ is the unit 2--sphere parametrized by $(\theta,\phi)$ and $\xi_r =\{x\,|\,0\leq x\leq r\}$, with $x=0$ marking a symmetry center
\footnote{Restrictions on $\DD$ occur if the $\T(t)$ have spherical topology or when there is a local curvature singularity. See the Appendix}. 
We define the ``quasi--local functional'' as the linear integral functional $\langle\hskip 0.1cm\rangle_*: X(\DD)\to \mathbb{R}$, such that for every scalar function $A\in X(\DD)$ we get the real number  
\begin{equation}\Aav_*[r] =  \frac{\int_0^r{A\,R^2R' {\rm d}x}}{\int_0^r{R^2R' {\rm d}x}}.\label{QLfunc}\end{equation}
where the notation ``$[r]$'' emphasizes the fact that the dependence on $r$ is that of a functional acting on functions in $X(\DD)$, to distinguish it from the dependence ``$(r)$'' in scalar function acting on $\DD$. In order to simplify notation, we use $\int_0^r{}\equiv\int_{x=0}^{x=r}{}$ to denote integration in the real interval $\xi_r$ and we will omit expressing explicitly the time dependence of $A,\,R,\,R'$ and $\Aav_*[r]$. Unless specified otherwise, we will assume inside an integral symbol that: \, $t$ is a constant parameter, all functions depend on $x$ and $R'=\partial R/\partial x$. A covariant definition equivalent to (\ref{QLfunc}) will be given in section \ref{covariant}. Meanwhile, we take this expression as an operative definition.\\

\noindent
{\underline{Definition 2. Quasi--local scalars.}} We can use the functional (\ref{QLfunc}) to construct scalar functions in $\DD$ with the same ``correspondence rule''. Consider the map $\JJ_*:X(\DD)\to X(\DD)$, such that for every $A\in X(\DD)$ and for all $r_0 \in\xi_r$
\begin{equation} A_*\equiv\JJ_*(A)\,:\DD\to\mathbb{R},\qquad A_*(r_0)=\Aav_*[r_0].\label{QLscals}\end{equation}
The real valued functions $A_*$ will be denoted ``quasi--local scalars'', as they generalize to any $A\in X$ the integral definition of the quasi--local mass--energy function (\ref{QLM_mu}) constructed with $\mu$.\\

\noindent
{\underline{Basic properties.}} The scalar functions $A_*(r)$, comply with the following useful properties:
\bse\label{props}\ba A - A_* &=& \frac{1}{R^3(r)}\int_0^r{A' R^3 {\rm d}x},\label{prop1}\\
    A_*{}' &=& \frac{3R'}{R}\,[A-A_*],\label{prop2}\\
\dot A_* &=& (\dot A)_*+3(\HH A)_*-3\HH_* A_*,\label{prop3}\ea\ese 
The functionals satisfy exactly the same properties above, with $A_*$ replaced by $\Aav_*[r]$ and taking $[r]$ as an arbitrary parameter. In order to simplify notation, we have adopted the following conventions:
\ba \dot A_* &\equiv& \left[\JJ_*(A)\right]\,\dot{}\ne \JJ_*(\dot A)=(\dot  A)_*,\nonumber\\
 A_*{}' &\equiv& \left[\JJ_*(A)\right]\,' \ne \JJ_*(A') = (A')_*,\nonumber\\
 (AB)_* &\equiv& \JJ_*(AB).\nonumber\\\label{notcon}\ea
It is straightforward to verify that these properties are perfectly self--consistent. \\

\noindent
{\underline{Functional vs function inside integrals}} The quasi--local functional and scalars (\ref{QLfunc}) and (\ref{QLscals}) behave differently under integration. The functions $A_*$ integrate as any scalar function, hence if we apply $\langle\hskip 0.1cm\rangle_*$ to $A_*$ we get:
\begin{equation} \langle A_*\rangle_*[r]=\frac{\int_0^r{A_*(x)R^2R'{\rm d}x}}{\int_0^r{R^2R'{\rm d}x}}\ne A_*(r),\label{non_ite_rule}\end{equation}
whereas, $\Aav_*[r]$ is just a real number that depends on the domain of a previous integral (\ref{QLfunc}), hence it complies with the following iteration rule:
\begin{equation} \langle \Aav_*[r]\rangle_*[r]=\frac{\int_0^r{\Aav_*[r] R^2R'{\rm d}x}}{\int_0^r{R^2R'{\rm d}x}}= \Aav_*[r],\label{ite_rule}\end{equation}
which allows us to regard the functional $\Aav_*[r]$ (but not the function $A_*(r)$) as an integral average distribution on $A$. See section \ref{QLaverage}.\\ 

\noindent
{\underline{Basic quasi--local scalars.}} Considering (\ref{QLM_mu}) and the definitions (\ref{QLfunc}) and (\ref{QLscals}), the quasi--local matter--energy energy $\mu_*=\JJ_*(\mu)$ is related to the quasi--local mass--energy function by
\begin{equation}\frac{\kappa}{3}\,\mu_* =\frac{2\MM}{R^3}.\label{QLM_mu2}\end{equation}
From (\ref{theta1}), (\ref{RR1}) and evaluating (\ref{QLfunc})--(\ref{QLscals}) for the scalars $\Theta,\,\RR$, their quasi--local equivalents of are
\bse\label{HKql}\ba \Theta_* &=& \JJ_*(\Theta)=\frac{3\dot R}{R},\label{theta2}\\
 \RR_* &=& \JJ_*(\RR)=\frac{6\,K}{R^2}\quad\Rightarrow\quad \frac{\dRR_*}{\RR_*}=-\frac{2}{3}\Theta_*,\label{RR2}
 \ea\ese 
In the following sections, until section \ref{covariant}, we will deal only with quasi--local scalar functions $A_*$. 

\end{quote}

\noindent
By applying (\ref{QLfunc})--(\ref{QLscals}) to (\ref{mu1})--(\ref{p1}), we get
\bse\ba \kappa\int_0^r{\mu R^2R'{\rm d}x} &=& \frac{\kappa}{3}\mu_*R^3=R\left(\dot R^2+K\right),\\
\kappa\int_0^r{p R^2R'{\rm d}x} &=& \frac{\kappa}{3}p_*R^3=-\frac{1}{3}\left[R\left(\dot R^2+K\right)+2\ddot R R\right],\ea\ese
which, with the help from (\ref{theta2})--(\ref{RR2}), lead to equations for $\mu_*,\,p_*,\,\Theta_*$ and $\RR_*$ that are identical to the Raychaudhuri and Friedman equation of FLRW models:
\ba \dot\Theta_* = -\frac{\Theta_*^2}{3}-\frac{\kappa}{2}\,(\mu_*+3p_*),\label{Raych2}\\
\left(\frac{\Theta_*}{3}\right)^2  = \frac{\kappa}{3}\,\mu_*-\frac{\RR_*}{6},\label{cHam2}\ea
where $\Theta_*^2 =[\JJ_*(\Theta)]^2 \ne \JJ_*(\Theta^2)$. These equations can be combined to yield the same FLRW energy conservation equation
\begin{equation}  \dot\mu_* = -(\mu_*+p_*)\,\Theta_*.\label{Econs2}\end{equation}
Applying (\ref{QLfunc})--(\ref{QLscals}) to (\ref{PP1}), (\ref{Sigma1}) and (\ref{EE_1}) yields the remaining scalars, $\Sigma$,\,$\PP$ and $\EE$, as deviations or fluctuations of $\Theta,\,\mu$ and $p$ from their quasi--local equivalents
\bse\label{PSE2}\ba \Sigma &=& -\frac{1}{3}[\Theta-\Theta_*],\label{Sigma_2}\\
\PP &=& \frac{1}{2}\,[p-p_*],\label{PP_2}\\
\EE &=& -\frac{\kappa}{6}\,\left[\mu-\mu_* +\frac{3}{2}(p-p_*)\right],\label{EE_2}\ea\ese
while the time derivative of the quasi--local mass--energy complies with
\begin{equation} 2\dot\MM = -\kappa \dot p_* R^2\dot R. \label{QLM3}\end{equation}
Since $\MM$ and $R$ are invariants of spherically symmetric spacetimes, it is evident that $\mu_*,\,p_*$ and $\Theta_*$ are covariant scalars, related to these invariants and their derivatives by means of (\ref{QLM_mu2}), (\ref{theta2}) and (\ref{QLM3}).  

\section{Evolution equations for the quasi--local scalars.}\label{eveqs}

A very convenient relation between the scalars $A$ and $A_*$ follows by defining the ``relative deviations'' or ``perturbations''
\begin{equation} \Da \equiv \frac{A-A_*}{A_*},\quad \Rightarrow\quad A = A_*\,\left[1+\Da\right].\label{Da_def}\end{equation}
which leads to the alternative covariant scalar representation of LTB spacetimes given by
\begin{equation} \{\Theta_*,\,\mu_*,\,p_*,\,\RR_*,\,\Dth,\,\Dm,\,\Dp,\,\DRR\}.\label{ql_scals}\end{equation}
We will call (\ref{ql_scals}) the ``quasi--local'' scalar representation and remark that it is fully equivalent to the local representation (\ref{loc_scals}). Regularity conditions for the $\Da$ functions are discussed in the Appendix.   

The ``1+3'' system (\ref{ev_theta_13})--(\ref{cHam_13}) is the set of evolution and constraint equations for (\ref{loc_scals}), we derive now the evolution and constraint equations for the representation (\ref{ql_scals}).    From (\ref{prop2}), the radial gradients of $\mu_*,\,p_*$ and $\HH_*$ can be given in terms of the $\delta$ functions by
\begin{equation} \frac{\Theta_*{}'}{\Theta_*} = \frac{3R'}{R}\,\Dth,\qquad \frac{\mu_*{}'}{\mu_*} = \frac{3R'}{R}\Dm,\qquad \frac{p_*{}'}{p_*} = \frac{3R'}{R}\Dp, \label{rad_grads}\end{equation}
while (\ref{Raych2}) and (\ref{Econs2}) are evolution equations for $\dot\mu_*$ and $\dot\Theta_*$. Hence, the evolution equations for $\Dm$ and $\Dth$ follow from the consistency condition
\begin{equation} \left[A_*{}'\right]\,\dot{}=\left[\dot A_*\right]',\end{equation}
applied to (\ref{Raych2}), (\ref{Econs2}) and (\ref{rad_grads}) for $A=\Theta_*$ and $\mu_*$. The result is the following set of autonomous evolution equations for the quasi--local scalar representation (\ref{ql_scals}):
\bse\label{eveqs_ql}\ba 
\fl\dot\mu_* &=& -\left[\,1+w\,\right]\,\mu_*\,\Theta_*,\label{evmu_ql}\\
\fl\dot\Theta_* &=& -\frac{\Theta_*^2}{3} -\frac{\kappa}{2}\,\left[\,1+3\,w\,\right]\,\mu_*,
\label{evHH_ql}\\
\fl\dDm &=& \Theta_*\,\left[\left(\Dm-\Dp\right)\,w-\left(1+w+\Dm\right)\Dth\right],
\label{evDmu_ql}\\
\fl\dDth &=& -\frac{\Theta_*}{3}\,\left(1+\Dth\right)\,\Dth + \frac{\kappa\mu_*}{6\,(\Theta_*/3)}\left[\Dth-\Dm+3w\,\left(\Dth-\Dp\right)\right],
\label{evDth_ql}
\ea\ese
where
\begin{equation}w\equiv \frac{p_*}{\mu_*}.\label{wdef}\end{equation}
The constraints associated with these evolution equations are simply the spatial  gradients (\ref{rad_grads}), while the Friedman equation (or Hamiltonian constraint) is (\ref{cHam2}). Notice that the constraints (\ref{rad_grads}) follow directly from differentiating the integral definition (\ref{QLfunc}), so by using the quasi--local variables we do not need to solve these constraints in order to integrate (\ref{evmu_ql})--(\ref{evDth_ql}).    

With the help of (\ref{Sigma_2})--(\ref{EE_2}) and (\ref{Da_def}), it is straightforwards to  prove that the evolution equations (\ref{evmu_ql})--(\ref{evDth_ql}) and the constraints (\ref{cHam2}) and (\ref{rad_grads}) are wholly equivalent to the 1+3 evolution equations (\ref{ev_theta_13})--(\ref{ev_EE_13}) and their constraints (\ref{cPP_13})--(\ref{cEE_13}) and (\ref{cHam_13}) for the local scalar representation (\ref{loc_scals}). Hence, they are also equivalent to the field and conservation equations. The system (\ref{evmu_ql})--(\ref{evDth_ql}) becomes fully determined once an ``equation of state'' that fixes $p_*$ as a function of $\mu_*$ is selected (see section \ref{EOS}). 

\section{A non--linear perturbation scheme}\label{perturb}

The dimensionless functions $\Da$ in (\ref{Da_def}) allow us to express the local scalars $A$ as non--linear relative fluctuations or perturbations of their quasi--local equivalents $A_*$. Notice, from (\ref{theta2})--(\ref{RR2}) and (\ref{Raych2})--(\ref{Econs2}), that the scalars $\{\Theta_*,\,\mu_*,\,p_*,\,\RR_*\}$ satisfy FLRW evolution laws. However, these are just the mappings under (\ref{QLscals}) of the LTB covariant scalars $\{\Theta,\,\mu_,\,p,\,\RR\}$, whose FLRW equivalents are also covariant scalars. This suggests that the role of the $\Da$ as spherical perturbations on a formal FLRW ``background'' state given by the $A_*$ can be rigorously justified by means of the known gauge invariant and/or covariant perturbation formalisms developed by Bardeen~\cite{bardeen} and Ellis, Bruni and Dunsbury~\cite{ellisbruni89,BDE,1plus3}.

A perturbation formalism (not necessarily linear) between an idealized spacetime, $\bar S$ (a FLRW cosmology), and an ``lumpy'' universe model, $S$, follows by defining a ``background model'' in $S$, constructed by objects (in $S$) that are the images of a suitable map $\Phi$ between objects in $\bar S$ to objects in $S$~\cite{ellisbruni89}. Though, perturbations in general are not uniquely defined by such an abstract map $\Phi$, because of the ambiguity of the gauge freedom associated with choices of coordinates and hypersurfaces. This fact requires carefully defining perturbed variables that are ``gauge--invariant''~\cite{bardeen}. However, in the specific case of setting a perturbation formalism when $S$ is an LTB spacetime we face a simpler task, as both the idealized FLRW and the perturbed LTB spacetimes can be completely described by scalar functions. Hence, $\Phi$ can be a map of scalars to scalars, while the fact that both spacetimes are spherically symmetric and are given in the same normal geodesic coordinate (or frame) representation, greatly suppresses most of the gauge freedom that we would find for a general $S$.  

For the FLRW-LTB case under consideration, the map $\Phi$ can be defined rigorously as follows. Let $\bar X$ and $X$ be, respectively, the sets of smooth integrable scalar functions in $\bar S$ and $S$, then for all covariant FLRW  scalars $\bar A\in \bar X$ (we denote FLRW objects with an over--bar) the map
\begin{equation} \Phi: \bar X\to X,\qquad \Phi(\bar A)=\JJ_*(A)=A_*\in X,\label{Phi}\end{equation}
defines a ``background model'' (associated to a FLRW cosmology) in LTB spacetimes through the quasi--local scalars $A_*$ (which are LTB objects satisfying FLRW dynamics). Given their common normal geodesic representation and considering the perturbations
\begin{equation} \Da = \frac{A-\Phi(\bar A)}{\Phi(\bar A)}\label{DPhi}\end{equation} 
associated with (\ref{Phi}), LTB spacetimes in the metric (\ref{LTB}) with source (\ref{Tab}) can be considered then as spherical non--linear ``perturbed'' FLRW cosmologies in the  ``synchronous'' gauge.   

The perturbation scheme that we described above can also be understood in terms of the gauge invariant and covariant (GIC) approach of Dunsbury, Ellis and Bruni~\cite{ellisbruni89,BDE}. Following these authors, a perturbation scheme on FLRW cosmologies is covariant if the ``lumpy'' model $S$ is described by variables defined in the framework of the ``1+3'' fluid flow variables $\mu,\,p,\,\Theta$ of (\ref{ev_theta_13})--(\ref{ev_EE_13}). Although our description of LTB spacetimes is not based on these scalars, it is still a covariant description because $\mu_*,\,p_*$ and $\Theta_*$, are covariant scalars by virtue of their connection with the invariants $\MM,\,R$ and their derivatives in (\ref{QLM_mu2}),  (\ref{theta2}) and (\ref{QLM3}). Hence, the formalism is covariant. 

As commented by Ellis and Bruni~\cite{ellisbruni89}, by virtue of the Stewart--Walker gauge invariance lemma \cite{SWlemma}, all covariant objects in $S$ that would vanish in the background $\bar S$ (a FLRW cosmology in this case) are gauge invariant (GI), to all orders, and also in the usual sense (as in \cite{bardeen}). The basic tensorial quantities in LTB spacetimes that vanish for a FLRW cosmology are the tensors $\Pi^{ab},\,\sigma^{ab}$ and $E^{ab}$ given by (\ref{Tab}), (\ref{sigma1}) and (\ref{EEtens}), which from (\ref{PPsc})--(\ref{EEsc}) and (\ref{Sigma_2})--(\ref{EE_2}), can be written as
\bse\label{PSE3}\ba \sigma^{ab} &=& - \frac{1}{3}\left[\Theta-\Theta_*\right]\,\Xi^{ab},\label{sigma_3}\\
 E^{ab} &=& -\frac{\kappa}{6}\,\left[\mu-\mu_*+\frac{3}{2}(p-p_*)\right]\,\Xi^{ab},\label{EE_3}\\
 \Pi^{ab} &=& \frac{1}{2}\,\left[p-p_*\right]\,\Xi^{ab},\label{PP_3}\ea\ese
where $\Xi^{ab}$ is the tensor defined in (\ref{PPsc})--(\ref{EEsc}), which is (for spherical symmetry) a covariant object. From (\ref{PPsc})--(\ref{EEsc}), (\ref{prop2}), (\ref{rad_grads}) and (\ref{sigma_3})--(\ref{PP_3}), it is evident that $\mu-\mu_*$,\, $p-p_*$ and $\Theta-\Theta_*$ and the radial gradients $\mu_*',\,p_*'$ and $\Theta_*'$  are all ``first order'' quantities (in the perturbation scheme) that are GI to all orders. Hence, from (\ref{Da_def}), the perturbation variables $\Dm,\,\Dp$ and $\Dth$ are also GI to all orders, though  $\mu_*,\,p_*,\,\Theta_*$ are not (which is expected because these are ``zero order'' background variables). Hence, the fluid flow dynamics of LTB spacetimes in the quasi--local scalar representation (\ref{ql_scals}) is that of spherical, non--linear GIC perturbations on a FLRW background. In section \ref{linear} we examine the connection with linear perturbations.              

It is worth remarking that $\RR_*$ does not appear in the dynamic equations (\ref{evmu_ql})--(\ref{evDth_ql}), though it can appear when specifying initial conditions to integrate these equations. This variable is GI only if the FLRW cosmology $\bar S$ is spatially flat, but its associated variables $\DRR$ and $\RR_*'$ are GI. In fact, from the Friedman equation (\ref{cHam2}) and (\ref{rad_grads}) we obtain
\bse\label{KdK}\ba \frac{\RR_*}{6} =  \frac{\kappa}{3}\mu_* -\left(\frac{\Theta_*}{3}\right)^2,\label{KmH}\\
\frac{\RR_*}{6}\,\DRR = \frac{\kappa}{3}\mu_*\Dm-2\left(\frac{\Theta_*}{3}\right)^2\Dth,\label{DKDmDH}\ea\ese
so that we can always eliminate either one of $\mu_*,\,\Theta_*$ or $\Dm,\,\Dth$ in terms of $\RR_*$ and $\DRR$, and construct a system of evolution equations equivalent to (\ref{evmu_ql})--(\ref{evDth_ql}), but describing the dynamics in terms of spatial curvature  $\RR_*$ and its perturbation $\DRR$.

\section{Equations of state.}\label{EOS}

The system (\ref{evmu_ql})--(\ref{evDth_ql}) is still undetermined, as there are no evolution equations for $p_*$ and $\Dp$. Equations (\ref{evmu_ql}) and (\ref{evHH_ql}) are formally identical to FLRW equations, and so are the evolution equations for the background, while (\ref{evDmu_ql}) and (\ref{evDth_ql}) are the evolution equations for the perturbations and thus convey the effects of inhomogeneity. As in any perturbative approach, we need to impose an ``equation of state'' (EOS) between $p_*$ and $\mu_*$ to determine the background subsystem (\ref{evmu_ql})--(\ref{evHH_ql}). Such a choice of an EOS also determines the perturbation equations.

Consider a commonly used ``barotropic'' equation of the form
\begin{equation} p_* = p_*(\mu_*).\label{BEOS}\end{equation} 
From (\ref{rad_grads}), we obtain the corresponding relation between fluctuations and perturbations of $\mu$ and $p$
\bse\label{BEOS_perflu}\ba p-p_* &=& \frac{d p_*}{d \mu_*}\,[\,\mu-\mu_*\,],\label{BEOSfluc}\\
\Dp &=& \frac{d\ln p_*}{d\ln\mu_*}\,\Dm.\label{BEOSpert}\ea\ese
An important and simple particular case of (\ref{BEOS}) is given by the linear EOS
\begin{equation} p_* = w_0\,\mu_*\,,\label{BEOSlin}\end{equation} 
where $w_0=w_0(t)$, and in particular $w_0$ can be a constant (so that $w_0=0$ is dust). In this case we have
\begin{equation}p-p_* = w_0\,[\,\mu-\mu_*\,],\qquad
\Dp = \Dm.\label{BEOSlinFP}\end{equation}
Notice that, in general, if an EOS like (\ref{BEOS}) is imposed it will not be satisfied by $p$ and $\mu$. But this is to be expected, since in a perturbation scheme fluctuations or perturbations do not satisfy, in general, the EOS of the background (not even in linear perturbations). The exception is the linear EOS (\ref{BEOS}), which implies $p=w_0\,\mu$, even if $w_0=w_0(t)$ because the quasi--local functional (\ref{QLfunc}) involves integrals in a hypersurface of constant $t$, hence $\JJ_*(w_0(t)\,\mu)=w_0(t)\,\JJ_*(\mu)$ for all $t$.  

It could be argued that, even if a given (\ref{BEOS}) is a physically reasonable EOS, it could yield a wholly unphysical relation between $p$ and $\mu$ through (\ref{BEOSfluc}). While this possibility cannot be ruled out, it is something that must be tested by looking at the solutions of (\ref{evmu_ql})--(\ref{evDth_ql}) for the particular EOS. If such a situation arises, then the specific LTB spacetime associated with this EOS could fail to provide an adequate non--linear perturbative description for the FLRW cosmology with this particular EOS.
However, even if such a development could occur for a given EOS, there is no reason to expect in self--gravitating systems that an EOS between $\mu_*$ and $p_*$ should also hold between local variables $p$ and $\mu$. This expectation is only correct in homogeneous conditions or for hydrodynamical systems characterized by short range interactions, much smaller than macroscopic scales~\cite{Padma_GTD}. Therefore, there is no reason to demand or to force such behavior of the EOS if we apply our formalism to self--gravitating relativistic sources of cosmological interest, like dark energy or the Chaplygin gas, whose fundamental interactions are still unknown and cannot be assumed to be of a hydrodynamical nature. 

Evidently, the plausibility of the non--linear perturbation formalism must be tested and judged for different EOS according to its predictions. In this context, it is worth remarking that quasi--local variables convey non--local information because they are integral quantities that depend on the behavior of the local variables in the whole integration range $\DD=\mathbb{S}^2\times\xi_r$. Since gravitation is a long range interaction, this non--local information can be relevant in the dynamical description of self--gravitating systems. We illustrate this fact in the following section using as example the classical ideal gas. Also, it has been suggested \cite{BRA4,wilt1,wilt2} that ``back--reaction'' is related to dynamical and observational effects of quasi--local energy. We discuss this issue in section \ref{BR}.  

\section{Quasi--local variables in the classical ideal gas.}\label{idealgas}

In order to illustrate directly the perturbation formalism under a  physically motivated and familiar EOS, we consider the classical ideal gas. This example illustrates the type of effects that could arise if we apply this scheme to sources like dark matter or dark energy. Following the arguments of the previous section, instead of (\ref{BEOS}) we choose the classical ideal gas EOS for background variables $\mu_*$ and $p_*$ 
\begin{equation}\mu_* = mc^2\,n_*+\frac{3}{2}n_*\, k\,T_*,\qquad p_* = n_*\,k\,T_*.
\label{IG_ave}\end{equation}
where $n_*=\JJ_*(n)$ and $T_*=\JJ_*(T)$ are,  respectively, the quasi--local scalar functions associated by means of (\ref{QLscals}) to particle number density and temperature, and $k$ is Boltzmann's constant. The EOS (\ref{IG_ave}) leads to a relation between the perturbations $\Dm,\,\Dp$ and $\delta^{(n)},\,\delta^{(T)}$, so that $n,\,T$ relate to $n_*,\,T_*$ by means of (\ref{Da_def}). The choice of background EOS (\ref{IG_ave}) transforms (\ref{evmu_ql})--(\ref{evDth_ql}) into a set of evolution equations for $n_*,\,T_*,\,\delta^{(n)},\,\delta^{(T)},\,\Theta_*$ and $\Dth$, leading to a description of this source as an inhomogeneous self--gravitating system in the non--linear perturbation formalism under consideration.

It is important to mention that the same EOS (\ref{IG_ave}), but relating local quantities $n$ and $T$, is strictly valid for non--interacting particles, or as an approximation to hydrodynamical sources under short range interactions (see \cite{hydro} for a hydrodynamical approach to the ideal gas by means of LTB spacetimes). For a self--gravitating system, given in terms of non--linear perturbations from an ideal gas in a FLRW background, we do not expect that $\mu$ and $p$ will be related by (\ref{IG_ave}).  

From (\ref{rad_grads}), (\ref{Da_def}) and (\ref{IG_ave}) we obtain the perturbations of $\mu$ and $p$ in terms of perturbations of $n$ and $T$
\bse\label{deltas_IG}\ba \Dm &=& \delta^{(n)}+\frac{\frac{3}{2}\,k\,T_*}{mc^2
+\frac{3}{2}\,k\,T_*}\,\delta^{(T)},\label{deltas_IG1}\\ \Dp &=& \delta^{(n)}+\delta^{(T)}.\label{deltas_IG2} \ea\ese
Using (\ref{Da_def}) and (\ref{deltas_IG1})--(\ref{deltas_IG2}) we can express $\mu$ and $p$ in terms  of the local variables $n,\,T$ and their fluctuations as
\bse\label{mup_local_IG}\ba 
\fl\mu &=& m c^2 n+\frac{3}{2}nkT\left[1-\frac{(n-n_*)\,(T-T_*)}{nT} \right]\nonumber\\
\fl &=& mc^2 n + \frac{3}{2}nkT\left[1 -\frac{\delta^{(n)}\delta^{(T)}}{(1+\delta^{(n)})(1+
\delta^{(T)})}\right],\label{mup_local_IG1}\\
\fl p &=& nkT\left[1-\frac{(n-n_*)\,(T-T_*)}{nT}\right]            
   = nkT\,\left[1 -\frac{\delta^{(n)}\delta^{(T)}}{(1+\delta^{(n)})(1+\delta^{(T)})}\right],\label{mup_local_IG2}
\ea\ese
which shows that (\ref{IG_ave}) leads to forms for the local pressure and internal energy [$(3/2)nkT$] that are equivalent to the ``standard'' (or ``non--perturbed'') EOS plus a virial correction that depends on the correlation between fluctuations of $n$ and $T$. This can also be appreciated by using the integral form (\ref{prop1}), leading to 
\ba \fl \frac{(n-n_*)\,(T-T_*)}{nT} &=&\frac{\int{n'R^3{\rm d}x}\int{T'R^3{\rm d}x}}{nTR^6}\nonumber\\
\fl &=& \frac{\int{\int{{\rm d}x {\rm d}\tilde x\, R^3(ct,x) R^3(ct,\tilde x)\,(\partial n/\partial x)\,
(\partial T/\partial \tilde x)}}}{nT R^6},\label{nTints}\ea
which is clearly a correlation density analogous to those found for Newtonian self--gravitational systems~\cite{saslaw1,saslaw2,saslaw3}. Although these articles considered gravitationally bound systems, and so their correlation terms will be different from those in a free--falling LTB ideal gas model, the qualitative comparison that we make with them is relevant, as it shows the effect of prescribing an EOS like (\ref{IG_ave}) on the relation between ``perturbed'' variables, $\mu$ and $p$, under a non--linear perturbation scheme. This perturbative treatment furnishes non--local information that adds correction terms to the standard ideal gas EOS ``$p=nkT$'', which is strictly valid for non--interacting particles. This extra information can be related to the long range interaction effects of gravity (and to ``back--reaction'', see section \ref{BR}) in scales where $(n-n_*)(T-T_*)/(nT)$ is not negligible. Since fluctuations like $(n-n_*)(T-T_*)$ are related to the gradients of $n_*$ and $T_*$ by (\ref{prop2}), the virial correction terms in (\ref{mup_local_IG1})--(\ref{mup_local_IG2}) are important when there is significant spatial variation in $n$ and $T$. 

\section{Connection with linear perturbations.} \label{linear}

Given the fact that $\Dm,\,\Dp,\,\Dth$ describe LTB spacetimes as spherical non--linear perturbations with respect to the ``background'' variables $\mu_*,\,p_*,\,\Theta_*$, it is important to examine the evolution equations (\ref{evmu_ql})--(\ref{evDth_ql}) in the linear regime, when these fluctuations are ``small'', {\it i.e.} $\Da\ll 1$. In these conditions we should be able to treat them (at least formally) as linear perturbations. 

In order to make this comparison, we derive a second order equation for the density perturbation
$\Dm$ for the special case of a barotropic EOS (\ref{BEOSlin}) with $w_0$ constant. By
differentiating both sides of (\ref{evmu_ql}) and using the remaining equations in (\ref{evmu_ql})--(\ref{evDth_ql}) to
eliminate all derivatives except $\ddot\delta^{(\mu)}$ and $\dDm$, we obtain
\ba \fl \ddot\delta^{(\mu)} &-&\frac{[\dDm]^2}{1+w_0+\Dm}+\frac{2\Theta_*}{3}\,
\dDm-\frac{\kappa}{2}\,(1+3w_0)\mu_*\,\Dm\,\left[1+w_0+\Dm\right]=0,\label{Dm2_eveq}\ea
which is an exact equation. For near homogeneous conditions, as assumed in linear perturbations,
(\ref{Dm2_eveq}) reduces to
\begin{equation}\ddot\delta^{(\mu)}+\frac{2\Theta_*}{3}\,
\dDm-\frac{\kappa}{2}\,(1+3w_0)(1+w_0)\mu_*\,\Dm=0, \label{lp_iso}\end{equation}
an equation that is formally identical to the evolution equation for linear density perturbations
of a fluid with this EOS around a FLRW background ($\mu_*,\,\Theta_*$) in the synchronous gauge~\cite{padma2}, and under the conditions of perturbation amplitudes much larger than the horizon \cite{padma2,hwangnoh}.  Notice that in the case of dust, $w_0=0$, we recover the well known perturbation equations in the comoving gauge (which for dust is a synchronous gauge as well) and without any restriction on the amplitude of the perturbations.

It is straightforward to show (with more complicated algebraic manipulations) that, for all EOS
among quasi--local variables, the evolution of $\Dm$ also reduces in the linear limit to
linear density perturbation equations in the synchronous gauge and in the long wavelength limit.
This is not surprising since the metric (\ref{LTB}) of LTB spacetimes is given in  coordinates
that are synchronous (because of its  geodesic 4--velocity) and comoving. There are known technical
difficulties with an equation like (\ref{lp_iso}) in this gauge (see Appendix of \cite{hwangnoh}). However, the
inconsistencies shown in this reference do not apply to the linear limit of LTB spacetimes with
$w\ne 0$ becuse the latter are also comoving and so ``$v_\alpha=0$'' from the outset in equations
(A1)--(A3) of \cite{hwangnoh}. Thus, the linear perturbation limit in the synchronous gauge with
long wavelengths is consistent. In a non--linear regime the correct equation is (\ref{Dm2_eveq}), not (\ref{lp_iso}), therefore the problems with this gauge in connection with the latter equation do not apply, as $\Da\ll 1$ is no longer valid.

\section{A covariant quasi--local functional.} \label{covariant}

A covariant expression for the quasi--local functional (\ref{QLfunc}) and scalar functions (\ref{QLscals}) follows rewriting these definitions in terms of integrals over the proper spatial volume element ${\rm d}\VV = \sqrt{\hbox{det}(h_{ab})}\,{\rm d}^3x$, where ${\rm d}^3x={\rm d}r\,{\rm d}\theta\,{\rm d}\phi$. Considering the same domain $\DD =\xi_r\times \mathbb{S}^2 \subset \T(t)$, with $\xi_r=\{x\,|\, 0\leq x \leq r\}$, the proper spatial volume for $\DD$ in LTB spacetimes is 
\bse\ba\VV &=& \int_\DD{{\rm d}\VV}=4\pi \int_{x=0}^{x=r}{\FF^{-1}\,R^2R' {\rm d}x},\label{LTB_propV}\\
 \FF &\equiv& [1-K]^{1/2},\label{F1}\ea\ese
where the $4\pi$ comes from the integration over $\mathbb{S}^2$. The definitions (\ref{QLfunc}) and (\ref{QLscals}) become ``weighed'' proper volume integrals:
\begin{equation}\langle A\rangle_*[r] = \frac{\int_\DD{A\,\FF\,{\rm d}\VV}}{\int_\DD{\FF\,{\rm d}\VV}}=\frac{\int_\DD{A\,{\rm d}\VV_*}}{\int_\DD{{\rm d}\VV_*}},\label{cQLfunc}\end{equation}
where 
\begin{equation} {\rm d}\VV_* \equiv \FF {\rm d}\VV,\qquad \VV_*=\int_\DD{\FF\,{\rm d}\VV}=\frac{4\pi}{3}R^3.\end{equation}
The ``weight factor'' (\ref{F1}) is a conserved invariant quantity ({\it i.e.} $\dot\FF=0$), which by virtue of (\ref{QLM_LTB}), can be given as
\begin{equation} \FF = \left[\dot R^2+\left(1-\frac{2\MM}{R}\right)\right]^{1/2},\label{F2}\end{equation}
Notice that $\FF$ is an invariant because $\dot R=u^a\nabla_a R$, and $\MM$ and $R$ are invariants in spherically symmetric spacetimes~\cite{MSQLM,szab,hayward1,hayward2}. As mentioned in section \ref{QLdefs}, $\MM$ can be interpreted as a total ``effective'' mass--energy in $\DD$. In the flat space limit $\MM/R\to 0$, this invariant reduces to the Lorentz contraction factor~\cite{misher}.  Also, if we replace $\MM$ in (\ref{F2}) with the ``Schwarzschild mass'' this equation is identical to the equation for radial geodesics in Schwarzschild geometry, with $\FF^2$ playing the role of the conserved quantity associated with the timelike Killing field (related to the binding energy).  

We show in the following section that $\FF$ is related to a binding energy integral. Another intuitive interpretation for $\FF$ follows from the Newtonian limit~\cite{hayward1}. The leading term of $\MM$ in this limit is the Newtonian rest mass integral: $\MM\approx (G/c^2)M_{\rm{newt}}$, while $\dot R=(1/c)\partial R/\partial t\ll 1$ and $\MM/R\approx GM_{\rm{newt}}/(c^2 R) \ll 1$, hence (\ref{F2}) becomes at first order
\begin{equation} \FF \approx 1+\frac{1}{c^2}\left[\left(\frac{\partial R}{\partial t}\right)^2-\frac{GM_{\rm{newt}}}{R}\right],\label{NlimF}\end{equation}
which is ``$\FF=1$'' plus a correction of order $c^{-2}$ given by the sum of Newtonian kinetic and gravitational potential energies per mass unit for comoving shells.  In a relativistic context this splitting of energy contributions is no longer valid.  

\section{Quasi--local vs proper averages.} \label{QLaverage}

The quasi--local functional, as expressed in covariant form by (\ref{cQLfunc}), can be considered a weighed average distribution with the conserved invariant $\FF$ playing the role of a formal ``probability density'' function. By its construction as a functional and its domain dependence, (\ref{cQLfunc}) complies with the properties of average distributions of random variables, such as the iteration rule: $\langle\Aav_*\rangle_*=\Aav_*$ given in (\ref{ite_rule}), as well as the definition of second order momenta (variance and covariance):
\bse\label{varcov}\ba \langle (A-\Aav_*)^2\rangle_* &=& \langle A^2\rangle_*-\Aav_*^2,
\label{variance}\\ \langle (A-\Aav_*)(B-\Bav_*\rangle_* &=& \langle AB\rangle_*-\Aav_*\Bav_*,
\label{covariance}\ea\ese  
where we have omitted the $[r]$, as the functional domain dependence is clear. The quasi--local averages constructed with (\ref{cQLfunc}) satisfy the mathematical properties (\ref{prop1})--(\ref{prop3}):
\bse\label{props_qlav}\ba A - \Aav_* &=& \frac{1}{\VV_*(r)}\int_0^r{A' \VV_* {\rm d}x},\label{prop_qlav1}\\
    \Aav_*{}' &=& \frac{\VV_*'}{\VV_*}\,[A-\Aav_*],\label{prop_qlav2}\\
\Aav_*\,\dot{} &=& \langle\dot A\rangle_*+3\langle\HH A\rangle_*-3\HHav_* \Aav_*,\label{prop_qlav3}\ea\ese 
where $\Aav_*\,\dot{}=\partial_0\Aav_*\ne\langle \partial_0 A\rangle_* =\langle\dot A\rangle_*$. It is evident from (\ref{non_ite_rule}) that quasi--local scalar functions $A_*$ do not satisfy (\ref{variance})--(\ref{covariance}), and so they are not ``averages'' of any sort.  

The quasi--local average is a generalization of the standard or proper (``weight'' $=1$) spatial average functional: $\langle\hskip 0.2cm\rangle:X(\DD)\to \mathbb{R}$, such that for every $A\in X(\DD)$ we get the real number
\begin{equation} \Aav[r]=\frac{\int_{\DD}{A\,{\rm d}\VV}}{\int_{\DD}{{\rm d}\VV}}=\frac{\int_0^r{A\,\FF^{-1}R^2R'\,{\rm d}x}}{\int_0^r{\FF^{-1}R^2R'\,{\rm d}x}}.\label{st_ave}\end{equation}
The average functionals $\langle\hskip0.1cm\rangle$ and $\langle\hskip0.1cm\rangle_*$ satisfy the same properties of average distributions of random variables, (\ref{ite_rule}) and (\ref{variance})--(\ref{covariance}). Also, the proper average satisfies (\ref{prop_qlav1})--(\ref{prop_qlav3}), with $\Aav_*$ and $\VV_*$ replaced by $\Aav$ and $\VV$. The form of the radial gradients of $\Aav$ in (\ref{prop_qlav2}) expressed in terms of metric functions is:
\begin{equation} \Aav' =\frac{\VV'}{\VV}\,\left[A-\Aav\right]=\frac{3R'}{R}\frac{\FFav}{\FF}\,\left[A-\Aav\right],\label{rad_grads_st}\end{equation}
where we applied the following relations: 
\bse\label{rel_aves}\ba \frac{\VV_*}{\VV} &=& \FFav,\label{rel_aves1}\\
 \Aav_* &=& \frac{\langle \FF A\rangle}{\FFav},\label{rel_aves2}\ea\ese
that follow directly from (\ref{cQLfunc}) and (\ref{st_ave}). 

Considering $\MM$ as the ``effective'' mass--energy in $\DD$ given by the integral of $\mu$ in (\ref{QLM1}), as discussed in section \ref{QLdefs}, suggests applying (\ref{cQLfunc}) and (\ref{st_ave}) to $\mu$. This leads to the following quasi--local and proper mass--energy functionals, related with the averages of $\mu$ and their corresponding volumes by 
\bse\label{mefs}\ba 2\MM[r] &=& \int_\DD{\mu \,{\rm d}\VV_*}=\frac{2G}{c^4}\muav_*\VV_*,\label{mefs1}\\
 2\MM_{\rm{pr}}[r] &=& \int_\DD{\mu\, {\rm d}\VV}=\frac{2G}{c^4}\muav\VV.\label{mefs2}\\
\ea\ese
As shown by these two expressions, in general, quasi--local (``effective'') and proper mass--energy is different and this difference (in spherical symmetry) can be used to define a ``binding energy'' \cite{MTW}, which in our case is the functional 
\begin{equation}\fl \B[r] =\frac{2G}{c^4}\left[\muav_*\VV_*-\muav\VV\right]= \frac{2G}{c^4}\int_{\DD}{[\FF-1]\,\mu\,{\rm d}\VV}=\frac{2G}{c^4}\,\langle (\FF-1) \mu\rangle\,\VV,\label{b_energ2}\end{equation}
where we used (\ref{rel_aves1})--(\ref{rel_aves2}). We have then an interpretation for $\FF$, as an invariant scalar function such that, for a given energy density $\mu$ and domain $\DD$, the sign of $\FF-1$ determines the sign of its corresponding binding energy $\B$ functional (\ref{b_energ2}).

Since pressure is also an energy density, similar mass--energy and binding energy functionals can be associated with $p$
\bse\label{mefsp}\ba \fl\W_*[r] &=& \frac{2G}{c^4}\int_{\DD}{p\,{\rm d}\VV_*}=\frac{2G}{c^4}\pav_*\VV_*,\label{mefsp1}\\
\fl \W_{\rm{pr}}[r] &=& \frac{2G}{c^4}\int_{\DD}{p\,{\rm d}\VV}=\frac{2G}{c^4}\pav\VV,\\
\fl\B_{(p)}[r] &=& \W_*-\W_{\rm{pr}}=\frac{2G}{c^4}\int_{\DD}{[\FF-1]\,p\,{\rm d}\VV}=\frac{2G}{c^4}\,\langle (\FF-1) p\rangle\,\VV,\label{b_energ2p}\ea\ese
Clearly, $\W_*$ and $\W_{\rm{pr}}$ can be interpreted as the proper and quasi--local work done by the averaged pressures $\pav_*$ and $\pav$ in the domain $\DD$. In fact, comparing (\ref{mefsp1}) with (\ref{QLM3}), we get: $2\dot\MM=-\W_*\dot\VV_*=-\W_*\VV_*\,\Thetaav_*$.

Regarding the averages $\Thetaav_*$ and $\Thetaav$, they are simply the proper time rate of change of $\VV_*$ and $\VV$:
\begin{equation} \Thetaav_*[r]=\frac{\dot \VV_*}{\VV_*} =\frac{3 \dot R}{R},\qquad \Thetaav[r]=\frac{\dot \VV}{\VV}=\frac{3\dot a}{a },\label{Thetaavs}\end{equation}
where we have introduced the ``scale factor'' $a\propto \VV^{1/3}[r]$ to characterize the (domain dependent) length scale of the averaging $\langle\hskip 0.1cm\rangle$, just as $R\propto \VV_*^{1/3}[r]$ (curvature radius or ``area--distance'' in $\DD$)  characterizes the domain dependent scale of $\langle\hskip 0.1cm\rangle_*$. Thus, for a dust source $R(r)\propto \rhoav_*^{1/3}[r]$ and $a(r)\propto \rhoav^{1/3}[r]$.

From (\ref{rel_aves1})--(\ref{rel_aves2})--(\ref{Thetaavs}), it is clear that the ratios
\begin{equation} \frac{\muav_*}{\muav}=\frac{\MM}{\MM_{\rm{pr}}}\FFav,\qquad \frac{\pav_*}{\pav}=\frac{\W}{\W_{\rm{pr}}}\FFav,\end{equation}
depend on the invariant $\FF$, and so, on the binding energy functional and on the spatial curvature through the function $K=(\RR_*\,R^2)/6$ in (\ref{RR2}). However, this is a non--local dependence, as it is determined by the behavior of the functions $\mu$ and $K$ in $\DD$. If we have positive, zero or negative spatial curvature in the whole $\DD$, then $\FF-1$ and $\B$ will be (respectively) negative, zero or positive. However, if a domain $\DD$ contains a local region with strong positive curvature (undergoing local collapse), binding energy can be still be positive if $\RR$ is asymptotically negative and $\DD$ sufficiently extended. Hence, as opposed to an FLRW cosmology, in an inhomogeneous LTB model the binding energy is not only domain dependent, but can change signs from one domain $\DD$ to another in each $\T$. If spatial curvature vanishes in $\DD$, then $\FF=1$ and $\langle\hskip 0.1cm\rangle_*=\langle\hskip 0.1cm\rangle$, binding energy is zero and the quasi--local averages become just proper averages.

As shown in \cite{BRA4,buchert,wilt1,wilt2}, the scale dependence of binding energies (associated with $\MM$ and $\MM_{\rm{pr}}$) are related to the so--called ``back--reaction'', and could provide important information connected with the theoretical interpretation of current cosmological observations. We examine this issue in the following section.

\section{Connection with ``Back--reaction''.} \label{BR}

``Back--reaction'' terms appear in a scalar averaging formalism, such as that developed by Buchert~\cite{buchert}, when the proper volume average functional (\ref{st_ave}) is applied to local scalar functions in fluid flow evolution equations like (\ref{ev_theta_13})--(\ref{ev_EE_13}). As we show in this section, the same terms arise in LTB spacetimes with both functionals (\ref{cQLfunc}) and (\ref{st_ave}). While for LTB spacetimes the fluid flow evolution equations, as well as (\ref{evmu_ql})--(\ref{evDth_ql}), are scalar covariant equations, for general spacetimes, the only covariant scalar equations are the Raychaudhuri, energy conservation and Friedman equations, (\ref{ev_theta_13}), (\ref{ev_mu_13}) and (\ref{cHam_13}). We rewrite these equations for LTB spacetimes as 
\bse\label{scal_evs_br}\ba \dot\Theta &=& -\frac{\Theta^2}{3}-\frac{\kappa}{2}\left[\mu+3p\right]-\frac{2}{3}\left(\Theta-\Theta_*\right)^2,\label{ev_theta_br}
\\
\dot\mu &=& -(\mu+p)\Theta+\left(\Theta-\Theta_*\right)\left(p-p_*\right),\label{ev_mu_br}
\\
\frac{\Theta^2}{9} &=& \frac{\kappa}{3}\mu-\frac{\RR}{6}-\frac{1}{9}\left(\Theta-\Theta_*\right)^2,\label{cHam_br}
\ea\ese
where we used (\ref{Sigma_2})--(\ref{EE_2}) to express $\Sigma$ and $\PP$ in terms of fluctuations of $\Theta$ and $p$ around their quasi--local equivalents $\Theta_*$ and $p_*$. 

Considering the derivative commutation property (\ref{prop_qlav3}) and (\ref{variance})--(\ref{covariance}), we apply the spatial average functional (\ref{st_ave}) to (\ref{ev_theta_br})--(\ref{cHam_br}) and work out the terms so that the averages $\Thetaav,\,\muav,\,\pav$ satisfy evolution equations that are FLRW plus ``corrections''. The result is
\bse\label{eqs_br}\ba \Thetaav\,\dot{} &=& -\frac{\Thetaav^2}{3}-\frac{\kappa}{2}\left[\,\muav+3\pav\,\right]+\QQ^{\rm{(k)}},\label{theta_br}\\
\muav\,\dot{} &=& -\left[\,\muav+\pav\,\right]\Thetaav-\QQ^{\rm{(d)}},\label{mu_br}\\
\frac{\Thetaav^2}{9} &=& \frac{\kappa}{3}\muav-\frac{1}{6}\left[\,\RRav+\QQ^{\rm{(k)}}\,\right],\label{Fr_br}\ea\ese 
where the ``corrections'' $\QQ^{\rm{(k)}}$ and $\QQ^{\rm{(d)}}$ are, respectively, the kinematic and dynamical ``back--reaction'' terms, given by
\ba \fl \QQ^{\rm{(k)}} &=& \frac{2}{3} \langle\C^{\rm{(k)}}\rangle,\qquad \C^{\rm{(k)}} \equiv  \left(\Theta-\Thetaav\,\right)^2-\left(\Theta-\Theta_*\right)^2,\label{Qk}\\
\fl \QQ^{\rm{(d)}} &=& \langle\C^{\rm{(d)}}\rangle,\qquad \C^{\rm{(d)}} \equiv  \left(\Theta-\Thetaav\,\right)\left(p-\pav\,\right)-\left(\Theta-\Theta_*\,\right)\left(p-p_*\right),\label{Qd}\ea
which coincide with the terms found elsewhere~\cite{buchert}. As a comparison, if we apply the quasi--local average functional (\ref{cQLfunc}) to (\ref{ev_theta_br})--(\ref{cHam_br}), we obtain evolution equations identical to (\ref{theta_br})--(\ref{Fr_br}), with $\langle\hskip 0.1cm\rangle_*$ replacing $\langle\hskip 0.1cm\rangle$, and with back--reaction terms given by
\ba \fl \QQ_*^{\rm{(k)}} &=& \frac{2}{3} \langle\C_*^{\rm{(k)}}\rangle_*,\qquad \C_*^{\rm{(k)}}\equiv  \left(\Theta-\Thetaav_*\,\right)^2-\left(\Theta-\Theta_*\right)^2,\label{Qkql}\\
\fl \QQ_*^{\rm{(d)}} &=& \langle\C_*^{\rm{(d)}}\rangle_*,\qquad C_*^{\rm{(d)}} \equiv  \left(\Theta-\Thetaav_*\,\right)\left(p-\pav_*\,\right)-\left(\Theta-\Theta_*\,\right)\left(p-p_*\right),\label{Qdql}\ea\\
which are exact analogues of (\ref{Qk}) and (\ref{Qd}) above. Because of the fact that $A_*(r)=\Aav_*[r]$ (from (\ref{QLscals})), the terms $\C_*^{\rm{(k)}}$ and $\C_*^{\rm{(d)}}$ in (\ref{Qkql}) and (\ref{Qdql}) vanish at $x=r$, but this would not make $\QQ_*^{\rm{(k)}}=0$ and $\QQ_*^{\rm{(d)}}=0$ because for $x<r$, both $\C_*^{\rm{(k)}}$ and $\C_*^{\rm{(d)}}$ are nonzero, and so the average $\langle\hskip 0.1cm\rangle_*$ would integrate over them (over $x$) yielding nonzero back--reaction terms. Only if spatial curvature is zero in all of $\DD$ we would have $\C_*^{\rm{(k)}}=\C_*^{\rm{(d)}}=0$ in all $\DD$, and then there would be no back--reaction~\cite{sussBR}. If using $\langle\hskip 0.1cm\rangle$, back--reaction also vanishes when spatial curvature is zero in all $\DD$, since then $\langle\hskip 0.1cm\rangle=\langle\hskip 0.1cm\rangle_*$.   

The basic motivation for using back--reaction terms in Buchert's formalism is to obtain an ``effective'' negative acceleration, $A_{\rm{eff}}$, that would mimic cosmic acceleration detected by observations. Considering a dust source:\, $p=\pav=p_*=\QQ^{\rm{(d)}}=0$ and $\mu=\rho c^2$, where $\rho$ is rest-mass density, the condition $A_{\rm{eff}}<0$ follows by rewriting (\ref{theta_br})--(\ref{Fr_br}) in terms of $A_{\rm{eff}}$, leading to the following condition (see \cite{buchert} for details)     
\begin{equation}A_{\rm{eff}}=\frac{\kappa}{2}\,\rhoav-\QQ^{\rm{(k)}} < 0,\label{efeacc}\end{equation}
where the kinematic back--reaction term, $\QQ^{\rm{(k)}}$, is given by (\ref{Qk}). 

As shown in \cite{sussBR}, a sufficient condition for $\QQ^{\rm{(k)}}\geq 0$ (and necessary for (\ref{efeacc})) is simply
\begin{equation} \C^{\rm{(k)}}\geq 0,\label{Cdef}\end{equation}
in the whole domain $\DD$. Clearly, (\ref{Cdef}) indicates that the necessary condition for (\ref{efeacc}) is basically a comparison between fluctuations of $\Theta$ with respect to its average, $\Thetaav$, with fluctuations with respect to its quasi--local equivalent scalar, $\Theta_*$ (but not the quasi--local average $\Thetaav_*$). In LTB spacetimes with nonzero pressure, we must add the ``dynamical'' back--reaction with the comparison of fluctuations $p-\pav$ and $p-p_*$ in (\ref{Qd}).  

It has been suggested by Wiltshire \cite{BRA4,wilt1,wilt2} that back--reaction, and so cosmic acceleration, is related to the spatial gradients of quasi--local energies between bounded structures that have decoupled from an expanding cosmic background. It is evident that this should be the case, since the fluctuations $\Theta-\Theta_*$ and $\Theta-\Thetaav$ that appear in (\ref{Qk}), (\ref{Qd}) and (\ref{Cdef}) are related to radial gradients of $\Theta_*$ and $\Thetaav$ by means of (\ref{rad_grads}) and (\ref{rad_grads_st}) 
\bse\ba \Theta'_* &=& \frac{\VV_*'}{\VV_*}\,[\Theta-\Theta_*]=\frac{3R'}{R}\,[\Theta-\Theta_*],\label{grad_Thql}\\
 \Thetaav' &=& \frac{\VV'}{\VV}[\Theta-\Thetaav]=\frac{3R'}{R}\,\frac{\FFav}{\FF}[\Theta-\Thetaav].\label{grad_Thav}\ea\ese
which translates into gradients of quasi--local energy, since (because of (\ref{theta2})) $\Theta_*=3\dot R/R$ can be identified with  a quasi--local ``kinetic'' energy density in the left hand side of Friedman equation (\ref{cHam2}), in its form (\ref{F2}), and in the Newtonian approximation (\ref{NlimF}), while $\Thetaav$ can be understood as an average ``kinetic'' energy density in this context. Since $\FF$ and $\FFav$ are directly related to the binding energy in (\ref{b_energ2}), we can relate these gradients to the bound states of observers in different domains $\DD$.  In fact, the back--reaction terms can be directly written as averages of gradients. Using (\ref{rad_grads}) and (\ref{rad_grads_st}), we rewrite (\ref{Qk}) and (\ref{Qd}) as a comparison of spatial gradients:
\ba \QQ^{\rm{(k)}}=\left\langle \left[\frac{\Thetaav'}{\VV'/\VV}\right]^2-\left[\frac{\Theta_*'}{\VV_*'/\VV_*}\right]^2\right\rangle,\\
\QQ^{\rm{(d)}}=\left\langle \frac{\Thetaav'\pav'}{\left[\VV'/\VV\right]^2}-\frac{\Theta_*'\,p_*'}{\left[\VV_*'/\VV_*\right]^2}\right\rangle,\ea
whose value, by virtue of (\ref{rel_aves1})--(\ref{rel_aves2}) and (\ref{b_energ2}),  depends on how $\FF$ determines the change of sign of the binding energy in the averaging domain $\DD$. These results are in agreement with and can give (model dependent) support to the arguments given by Wiltshire in \cite{BRA4,wilt1,wilt2}. The connection between quasi--local variables and functionals with the issue of back--reaction deserves a proper examination that is being undertaken in a separate article.  

 \section{Conclusion and possible avenues for future research.}\label{conclusion} 
 
We have examined a class of spherically symmetric inhomogeneous spacetimes (denoted  ``LTB spacetimes'') associated with the LTB metric (\ref{LTB}), whose source is an anisotropic fluid (\ref{Tab}). In general, LTB spacetimes can be compatible with a wide variety of sources and ``equations of state'', and in particular, we can always recover the well known LTB dust solutions by setting $p=0$. 

Following the theoretical formalism that we have presented in this article (which we summarized in the introduction), LTB spacetimes can be used as models of interacting dark matter and dark energy, as well as models to test alternative proposals based on back--reaction effects.  As non--linear perturbations of a FLRW cosmology in the quasi--local scalar representation, they are ideal theoretical tools to construct appealing models in cosmology that generalize (to idealized but non--trivial and non--linear inhomogeneity conditions) a large ammount of known solutions in modern cosmology that have been derived and examined in a FLRW context (or with linear perturbations). 

While the integration of the evolution equations (\ref{evmu_ql})--(\ref{evDth_ql}) definitely requires numerical work, these equations can be handled as a system of ODE's in which radial dependence enters as a parameter (see Appendix C of \cite{suss08}), thus the evolution system can be treated as a ``dynamical system'' using numerical techniques that are quite accessible and can be handled by known and widely used packages. In fact, the quasi--local scalar representation has already been used successfully to conduct a ``dynamical system'' study of LTB dust solutions.~\cite{suss08}         

The connection with back--reaction is straightforward. The ``kinematic'' and ``dynamical'' back--reaction terms that emerge from Buchert's averaging formalism are both expressible as differences between squared fluctuations (variances and covariances) of $\Theta$ and $p$ around their proper averages, $\Thetaav$ and $\pav$, and their quasi--local equivalents, $\Theta_*$ and $p_*$. These fluctuations can also be given in terms of spatial gradients of $\Thetaav,\,\pav$ and $\Theta_*,\,p_*$, and these expressions can also be related to the (domain dependent) binding energy integral $\B$. These comparisons provide a quantitative (though model dependent) framework to derive the conditions for the existence of back--reaction~\cite{sussBR}, and a subsequent negative ``effective'' acceleration that could mimic cosmic acceleration. This framework provides also a substantative appraisal to the the qualitative arguments of Wiltshire~\cite{BRA4,wilt1,wilt2} that characterize back--reaction as an effect related to the existence of spatial variation in quasi--local energy between bound observers and an expanding cosmic background. 

Evidently, the predictions of LTB spacetimes, as sources that are inhomogeneous at specific scales, must be tested in reference to observations \cite{InhObs1,InhObs2,InhObs3,InhObs4,InhObs5,InhObs6,InhObs7} (see also \cite{Mota, test_BR} and references quoted therein). However, at least in scales below the so--called homogeneity scale (100--300 Mpc), models built with LTB spacetimes can provide useful information about non--linear effects associated with inhomogeneity, such as ``spherical collapse'' models depicting gravitational collapse in local regions embedded in an expanding homogenous background, effects of dark energy in large scale dynamics of galactic superclusters, back--reaction terms in the dynamics of averaged scalars, quasi--local energies and non--local phenomena. In a follow up article we will apply LTB spacetimes as models for inhomogeneous dark energy sources, such as the Chaplygin gas, interactive mixtures of dark matter and dark energy and scalar fields. We will also re--examine in more depth, and in separate articles,  the connection between quasi--local variables and back--reaction (and the ``effective'' acceleration). 

It is important to stress that the quasi--local representation can be reformulated in terms of fluctuations, variances and covariances of quasi--local average distributions that can be compared to those associated with the ``standard'' proper volume average of Buchert's formalism. In particular, the quasi--local average introduces a ``weight function'' that is an invariant that could be interpreted as some sort of non--trivial probability density. Further developments of his framework could lead to a definition of a quasi--local ``gravitational entropy'' functional (in the context of \cite{grav_entr}), and/or to possible theoretical analogies with statistical mechanics and black hole thermodynamics. This is a very appealing possible avenue for future research and is presently under consideration.

Finally, the formalism that we have presented can be readily generalized beyond spherical symmetry to the ``quasi--spherical'' Szekeres models, which do not admite isometries~\cite{kras,krashela}. While the metric coefficients, as well as all covariant quantities, in these models depend on all coordinates, it has been proven that the proper spatial volume average of scalars is independent of the ``non--radial'' coordinates \cite{bolejko} (see also \cite{kasai}). As we prove in a forthcomming article, this is also the case with the quasi--local functions and averages. Thus, the techinques used in this article are valid for these general models with just minimal adjustments, but the latter models are no longer spherically symmetric and so provide a less idealized framework.  

\section*{Appendix: some regularity issues}

As mentioned in section IV, we have assumed that the integration domain $\DD=\xi_r\times \mathbb{S}^2$ with $\xi_r=\{x\,|\,0\leq x\leq r\}$, containing a symmetry center at $x=0$, is fully regular. In particular, if the hypersurfaces $\T(t)$ have spherical $\mathbb{S}^3$ topology, there is a second symmetry center, $x=r_c$, and so the domain is restricted to $x\leq r_c$.  Another restriction comes from the well known fact that curvature singularities  (initial and collapsing) arise in LTB dust models. In particular, there is a re--collapsing singularity in regions with positive spatial curvature, $K>0$, and when $\Theta<0$~\cite{ltbstuff,suss02}. Since this type of curvature singularities is bound to occur as well in the more general LTB models, it is worthwhile commenting on its possible effects.\\ 

The coordinate locii of initial or collapsing singularities in dust solutions are (in general) not simultaneous ({\it i.e.} not marked by a constant $t$). Even if parameters can be set up to have a simultaneous initial singularity~\cite{ltbstuff,suss02}, in general, the collapsing singularity in regions with $K>0$ is not simultaneous, but marked by a curve $[ct(r_{\rm{coll}}), r_{\rm{coll}}]$ in the $(ct,r)$ coordinate plane, so that $R(ct(r_{\rm{coll}}), r_{\rm{coll}})=0$, and curvature scalars diverge as comoving observers approach these coordinates (notice that this is different from a symmetry center, $r=r_c$, where $R(ct,r_c)=0$ regularly for all $t\ne t(r_{\rm{coll}})$). Thus, in any such collapsing region the hypersurfaces $\T(t)$ for $t\geq t(r_{\rm{coll}})$ are only regular for the semi open subset $\xi_{\rm{coll}}\equiv \{ x\,|\, r_{\rm{coll}} < x \leq r\} \subset \xi_r$. However, the existence of this (and the initial) singularity has no consequence in the definition of the quasi--local functional and scalar functions, because the involved integrals can be treated simply as standard improper integrals. We define at each $\T(t)$ the incumbent integrals with their lower integration limit as $y =r_{\rm{coll}}+\epsilon$, for an arbitrarily small $\epsilon>0$, and then obtain the limit as $\epsilon\to 0$. Off course, since $\Theta\to -\infty$ in this limit, $\Theta_*$  and $\Thetaav_*$ might diverge as well, but these functions are well defined in the range $\xi_{\rm{coll}}$.\\

Another important regularity issue, also present in dust LTB solutions, is the possibility of unphysical ``shell crossing'' singularities, which might occur if $R'=0$ for non--comoving coordinate values ({\it i.e.} coordinate values not equal to constant $r$). See \cite{ltbstuff,suss02} for a comprehensive discussion. It is important to distinguish the shell crossings from ``turning values'' at a given constant value of $r=r_{\rm{tv}}$ where $R'(r_{\rm{tv}})=0$ regularly, as in the case where the $\T(t)$ have spherical $\mathbb{S}^3$ topology (or for ``closed'' FLRW spatial sections). The possible effect of a shell crossing singularity in LTB spacetimes would be to make the functions $\Da$ ill defined. From (\ref{prop2}), (\ref{rad_grads}) and (\ref{prop_qlav2}), the $\Da$ will be regular if $R'=0$ implies $A_*'=0$, which will happen at $r=r_{\rm{tv}}$. But if $R'(ct,r)=0$ and $r\ne r_{\rm{tv}}$, then $R'=0$ will happen with $A_*'\ne 0$, and consequently, $\Da$ diverges. Since $A=A_*[1+\Da]$, this would mean that $A$ diverges for a finite $A_*$, a wholly unphysical and unacceptable situation. Therefore, a standard regularity condition in LTB dust solutions is to demand absence of shell crossings, or $R'>0$, save at turning values for hypersurfaces $\T$ with spherical ($\mathbb{S}^3$) or ``wormhole'' topologies  ($\mathbb{S}^2\times \mathbb{R}$ or $\mathbb{S}^2\times \mathbb{S}^1$) that have no symmetry centers. This regularity condition should also be demanded for the more general LTB spacetimes, which could exhibit spherical topology. For the ``wormhole'' topologies, the integrals (\ref{QLfunc}) and (\ref{cQLfunc}) defining the quasi--local functional could be unbounded, but this could be avoided if the lower integration limit is defined asymptotically (see \cite{suss08} for an example).

\end{document}